\documentclass[12pt]{article}
\usepackage{amssymb}
\usepackage{color}
\usepackage{indentfirst}
\usepackage{amsmath}
\usepackage{enumerate}
\usepackage{cite}

\newtheorem{theorem}{Theorem}
\newtheorem{lemma}{Lemma}

\newtheorem{example}{Example}

\def\qi#1 {\fbox {\footnote {\ }}\ \footnotetext { From Qi: {\color{red}#1}}}

\begin{document}
\title{Complete weight enumerators of a class of linear codes with four or five weights}
\author{Xina Zhang\\
College of Mathematics and Statistics, Northwest Normal \\University,
 Lanzhou,  Gansu 730070,  P.R. China\\
Email: zhangxina11@163.com}
\date{}
\maketitle

\begin{abstract}
In this paper, based on the theory of defining sets, a class of four-weight or five-weight linear codes over $\mathbb{F}_p$ is constructed. The complete weight enumerators of the linear codes are determined by means of Weil sums. In some case, there is an almost optimal code with respect to Griesmer bound, which is also an optimal one according to the online code table. This is an extension of the results raised by Zhang et al.(2020).

\textbf{Key words:} linear codes, complete weight enumerator, Weil sums, almost optimal codes

\end{abstract}

\section{Introduction}
In this study, $p$ is an odd prime and assume $q=p^e$ for a positive integer $e$. Let $\mathbb{F}_p$ and $\mathbb{F}_q$ denote the finite field with $p$ and $q$ elements, respectively. We denote by $Tr$ the absolute trace function \cite{LN} from $\mathbb{F}_q$ onto $\mathbb{F}_p$, and use $\mathbb{F}^{*}_q$ and $\mathbb{F}^{*}_p$ to denote the multiplicative group of $\mathbb{F}_q$ and $\mathbb{F}_p$. Obviously, $\mathbb{F}_{q}^{*}=\mathbb{F}_{q}\setminus\{0\}$, and $\mathbb{F}_{p}^{*}=\mathbb{F}_{p}\setminus\{0\}$.

An $[n,k,d]$ \emph{linear code} $\mathcal{C}$ over $\mathbb{F}_q$ is a $k$-dimensional subspace of
$\mathbb{F}_q^n$ with minimum Hamming distance $d$. Let $A_i$ be the number of codewords with Hamming weight $i$ in a code $\mathcal{C}$. The weight enumerator of $\mathcal{C}$ is defined by
$$1+A_1z+A_2z^2+\ldots+A_nz^n,$$
and the sequence $(1, A_1, \ldots, A_n)$ is called the \emph{weight distribution} of $\mathcal{C}$\cite{HP}. If $|\{1\leq i\leq n: A_i\neq 0\}|=t,$ then we say $\mathcal{C}$ a $t$-weight code.

Let us denote $\mathbb{F}_p=\{\omega_{0}=0, \omega_{1}, \ldots, \omega_{p-1}\}$. The complete weight enumerator $cwe[\mathbf{c}]$ of a codeword $\mathbf{c}=(c_{0}, c_{1}, \ldots, c_{n-1})\in \mathbb{F}^{n}_p$ is defined as $$cwe[\mathbf{c}]=\omega_{0}^{t_{0}}\omega_{1}^{t_{1}}\ldots\omega_{p-1}^{t_{p-1}} ,$$
where $t_{i}$ is the number of coordinates of $\mathbf{c}$ equal to $\omega_{i}$. Obviously, $\sum_{i=0}^{p-1}t_{i}=n .$ The \emph{complete weight enumerator}\cite{HP} of the code $\mathcal{C}$ is the polynomial $$\mathbf{CWE}(\mathcal{C})=\sum_{\mathbf{c}\in\mathcal{C}}cwe[\mathbf{c}].$$

The complete weight enumerator of linear codes over finite field $\mathbb{F}_p$ can show the frequency of each symbol appearing in each codeword. For binary linear codes, the complete weight enumerators are just the weight enumerators, while the weight enumerator of nonbinary linear codes can be easily obtained from the complete weight enumerators. The complete weight enumerator not only can be used to compute the deception probability of some authentication codes constructed from linear codes\cite{CTTX}, but also can be employed to calculate the Walsh transform of monomial functions over finite fields\cite{TA}. Though determining the complete weight enumerator is an interesting work, it is difficult in general. Thus, only a few research about specific codes involved this topic \cite{IK,AA,CQF,SCQ}.

One of the constructions of linear codes is based on a proper selection of a subset of finite fields\cite{DCS07}. That is, let $D = \{ {d_1},{d_2}, \ldots ,{d_n}\} \subseteq {F_q}$. A linear code of length $n$ over $\mathbb{F}_p$ is defined as
\begin{equation}
\mathcal{C}_D =\{(Tr(xd_1),Tr(xd_2), \ldots ,Tr(xd_n)):x \in \mathbb{F}_q\},
\end{equation}
the set $D$ is called the \emph{defining set} of linear code $\mathcal{C}_D$. Based on this method, literatures\cite{XS,QFKD}    constructed different linear codes and determines their complete weight enumerators and weight enumerators. Particularly, there are some optimal linear codes in the literatures\cite{GXSJ,SXC}.

Let $\mathcal{C}$ be an $[n,k,d]$ code over $\mathbb{F}_{q}$ with $k\geqslant1$, then the well-known \emph{Griesmer bound}\cite{HP} is given by
$$n\geqslant\sum_{i=0}^{k-1}\lceil\frac{d}{q^{i}}\rceil.$$

An $[n,k,d]$ code is called \emph{optimal} if no $[n,k,d+1]$ code exists, and is called \emph{almost optimal} if the $[n,k,d+1]$ code is optimal\cite{HY}.

By means of the construction method mentioned above, Zhang et al.\cite{XXRF} constructed a class of linear codes and presented their weight distributions, with the defining set
$D = \{(x_1, x_2) \in \mathbb{F}_q^2 : Tr(x_1^{{p^s} + 1}) = 1, Tr(x_2) = 1  \},$  where $p$ is an odd prime, $q=p^m,$ and $m=2s.$
In this paper, with the same means, we generalize the construction of the defining set, and obtain a class of linear codes with four or five weights, which include some almost optimal codes. And making use of Weil sums\cite{R,S,C}, we will determine not only weight distributions but also complete weight enumerators of these codes.


\section{Main Results}\label{theorem}
In this section, we present the main results, including the construction, the parameters, the complete weight enumerator and the weight distribution of the linear code $\mathcal{C}_D$. The proofs will be given in the following section.

We begin this section by selecting the definition set
\begin{equation}
D = \{(x_1, x_2) \in \mathbb{F}_q^2 : Tr(x_1^{{p^l} + 1}) = 1, Tr(x_2) = 1  \}
\end{equation}
to construct linear code
\begin{equation}
\mathcal{C}_D = \{(Tr(ax_1+bx_2)_{(x_1, x_2)\in D}) : a, b \in \mathbb{F}_q  \},
\end{equation}
where $l$ is a positive integer. Recall $q=p^e$, and let $s=\gcd(l,e)$ be the greatest common divisor of integers $l$ and $e$. The code $\mathcal{C}_{D}$ will be discussed under the assumption that $e/s$ is even with $e=2m$ and $m\geq1$. Then the weights, weight distributions and complete weight enumerators of the linear codes are studied by utilizing some results of Weil sums\cite{S,C}.

The following Theorems \ref{weight1}-\ref{weight2} are the main results of this paper.

\begin{theorem}\label{weight1}
Let $g$ be a generator of $\mathbb{F}_{p}^{*}$. If $ m/s \equiv 1\,mod\,2,$ then the weight distribution of the code $\mathcal{C}_D$ with the parameter $[p^{2e-2}+p^{e+m-2}, 2e]$ is listed in table \ref{1}. And the complete weight enumerator is
\begin{displaymath}
\begin{aligned}
  & \omega_{0}^{p^{2e-2}+p^{e+m-2}}+\sum_{\alpha=1}^{p-1}\omega_{g^{\alpha}}^{p^{2e-2}+p^{e+m-2}}
   +(p^{e-1}+p^{m-1})\sum_{\alpha=1}^{p-1}\omega_{0}^{p^{2e-3}}\prod_{\rho \in \mathbb{F}_{p}^{*}}\omega_{\rho}^{p^{2e-3}-p^{e+m-2}\cdot\eta(\rho^{2}-2g^{\alpha}\rho)}\\
  & +p^{e}(p^{e}-p)\prod_{\rho \in \mathbb{F}_{p}}\omega_{\rho}^{p^{2e-3}+p^{e+m-3}}
    +(p^{e-1}-p^m+p^{m-1}-1)\omega_{0}^{p^{2e-3}+p^{e+m-2}}\prod_{\rho \in \mathbb{F}_{p}^{*}}\omega_{\rho}^{p^{2e-3}}\\
  & +(p^{e-1}+p^{m-1})\sum_{\beta=1}^{\frac{p-1}{2}}\omega_{0}^{p^{2e-3}+p^{e+m-2}\cdot\eta(-1)}\prod_{\rho \in\mathbb{F}_{p}^{*}}\omega_{\rho}^{p^{2e-3}-p^{e+m-2}\cdot\eta(\rho^{2}-4g^{2\beta+1})}\\
  &+(p^{e-1}+p^{m-1})\sum_{\beta=1}^{\frac{p-1}{2}}\omega_{0}^{p^{2e-3}-p^{e+m-2}\cdot\eta(-1)}\omega_{2g^{\beta}}^{p^{2e-3}}
  \omega_{p-2g^{\beta}}^{p^{2e-3}}\prod_{\rho \in \mathbb{F}_{p}^{*} \atop \rho \neq  \pm2g^{\beta}}\omega_{\rho}^{p^{2e-3}-p^{e+m-2}\cdot\eta(\rho^{2}-4g^{2\beta}))}\\
  & +(p^{e-1}+p^{m-1})\sum_{\alpha=1}^{p-1}\sum_{\beta=1}^{\frac{p-1}{2}}
  \omega_{0}^{p^{2e-3}+p^{e+m-2}}\prod_{\rho \in \mathbb{F}_{p}^{*}}\omega_{\rho}^{p^{2e-3}-p^{e+m-2}\cdot\eta(\rho^{2}-2g^{\alpha}\rho+g^{2\beta+1})}\\
  & +(p^{e-1}+p^{m-1})\sum_{\substack{\alpha=1\\\alpha\neq\beta\\\alpha\neq\frac{p-1}{2}+\beta}}^{p-1}\sum_{\beta=1}^{\frac{p-1}{2}}
  \omega_{0}^{p^{2e-3}-p^{e+m-2}}\prod_{\rho \in \mathbb{F}_{p}^{*}}\omega_{\rho}^{p^{2e-3}-p^{e+m-2}\cdot\eta(\rho^{2}-2g^{\alpha}\rho+g^{2\beta})}\\
  & +(p^{e-1}-p^{m}+p^{m-1}-1)\sum_{\alpha=1}^{p-1}\omega_{g^{\alpha}}^{p^{2e-3}+p^{e+m-2}}\prod_{\rho \in \mathbb{F}_{p} \atop \rho \neq g^{\alpha}}\omega_{\rho}^{p^{2e-3}}
  \end{aligned}
\end{displaymath}
where $\eta$ is the quadratic character over $\mathbb{F}_{p}^{*}$, and is extended by $\eta(0)=0$. Obviously, the code is at most $5$-weight. When $m=1, s=1$, note that $p^{2e-2}+p^{e+m-2}=p^{2e-3}(p-1)+2p^{e+m-2}$, thus the code is a four-weight linear code.

\begin{table}
\begin{center}
\caption{The weight distribution of $\mathcal{C}_D$ when $ m/s \equiv 1\,mod\,2 $}\label{1}
\begin{tabular}{ll}
\hline\noalign{\smallskip}
Weight  &  Multiplicity   \\
\noalign{\smallskip}
\hline\noalign{\smallskip}
$0$  &  1 \\
$p^{2e-2}+p^{e+m-2}$  &  $ p-1 $    \\
$(p-1)(p^{2e-3}+p^{e+m-3})$  &  $p^e(p^e-p)$     \\
$p^{2e-3}(p-1)$  &  $\frac{1}{2}(2p^{m-1}-3p^{m}+p^{m+1}+2p^{e-1}-p^{e}+p^{e+1}-2)$     \\
$  p^{2e-3}(p-1)+2p^{e+m-2}$  &  $ \frac{1}{2}(2p^{m-1}-3p^m+p^{m+1}+2p^{e-1}-3p^e+p^{e+1})$     \\
$ p^{2e-3}(p-1)+p^{e+m-2}$ & $ 1-p-2p^{m-1}+3p^m-p^{m+1}-2p^{e-1}+2p^e $     \\
\noalign{\smallskip}
\hline
\end{tabular}
\end{center}
\end{table}
\end{theorem}

\begin{theorem}\label{weight2}
Let $g$ be a generator of $\mathbb{F}_{p}^{*}$. If $m \geq s+1$ and $ m/s \equiv 0\,mod\,2,$ then the weight distribution of the code $\mathcal{C}_D$ with the parameter $[p^{2e-2}+p^{e+m+s-2}, 2e]$ is listed in table \ref{2}. And the complete weight enumerator is
\begin{displaymath}
\begin{aligned}
  & \omega_{0}^{p^{2e-2}+p^{e+m+s-2}}+\sum_{\alpha=1}^{p-1}\omega_{g^{\alpha}}^{p^{2e-2}+p^{e+m+s-2}} +p^{e}(p^{e}-p^{1-2s})\prod_{\rho \in \mathbb{F}_{p}}\omega_{\rho}^{p^{2e-3}+p^{e+m+s-3}}\\
   &+(p^{e-2s-1}-p^{m-s}+p^{m-s-1}-1)\omega_{0}^{p^{2e-3}+p^{e+m+s-2}}\prod_{\rho \in \mathbb{F}_{p}^{*}}\omega_{\rho}^{p^{2e-3}}\\
   &+(p^{e-2s-1}+p^{m-s-1})\sum_{\alpha=1}^{p-1}\omega_{0}^{p^{2e-3}}\prod_{\rho \in \mathbb{F}_{p}^{*}}\omega_{\rho}^{p^{2e-3}-p^{e+m+s-2}\cdot\eta(\rho^{2}-2g^{\alpha}\rho)}\\
  &+(p^{e-2s-1}+p^{m-s-1})\sum_{\beta=1}^{\frac{p-1}{2}}\omega_{0}^{p^{2e-3}+p^{e+m+s-2}\cdot\eta(-1)}\prod_{\rho \in \mathbb{F}_{p}^{*}}\omega_{\rho}^{p^{2e-3}-p^{e+m+s-2}\cdot\eta(\rho^{2}-4g^{2\beta+1})}\\
  &+(p^{e-2s-1}+p^{m-s-1})\sum_{\beta=1}^{\frac{p-1}{2}}\omega_{0}^{p^{2e-3}-p^{e+m+s-2}\cdot\eta(-1)}\omega_{2g^{\beta}}^{p^{2e-3}}
  \omega_{p-2g^{\beta}}^{p^{2e-3}}\prod_{\rho \in \mathbb{F}_{p}^{*} \atop \rho \neq \pm2g^{\beta}}\omega_{\rho}^{p^{2e-3}-p^{e+m+s-2}\cdot\eta(\rho^{2}-4g^{2\beta}))}\\
  & +(p^{e-2s-1}+p^{m-s-1})\sum_{\alpha=1}^{p-1}\sum_{\beta=1}^{\frac{p-1}{2}}
  \omega_{0}^{p^{2e-3}+p^{e+m+s-2}}\prod_{\rho \in \mathbb{F}_{p}^{*}}\omega_{\rho}^{p^{2e-3}-p^{e+m+s-2}\cdot\eta(\rho^{2}-2g^{\alpha}\rho+g^{2\beta+1})}\\
  & +(p^{e-2s-1}+p^{m-s-1})\sum_{\substack{\alpha=1\\\alpha\neq\beta\\\alpha\neq\frac{p-1}{2}+\beta}}^{p-1}\sum_{\beta=1}^{\frac{p-1}{2}}
  \omega_{0}^{p^{2e-3}-p^{e+m+s-2}}\prod_{\rho \in \mathbb{F}_{p}^{*}}\omega_{\rho}^{p^{2e-3}-p^{e+m+s-2}\cdot\eta(\rho^{2}-2g^{\alpha}\rho+g^{2\beta})}\\
  & +(p^{e-2s-1}-p^{m-s}+p^{m-s-1}-1)\sum_{\alpha=1}^{p-1}\omega_{g^{\alpha}}^{p^{2e-3}+p^{e+m+s-2}}\prod_{\rho \in \mathbb{F}_{p} \atop \rho \neq g^{\alpha}}\omega_{\rho}^{p^{2e-3}}
  \end{aligned}
\end{displaymath}
where $\eta$ is the quadratic character over $\mathbb{F}_{p}^{*}$, and is extended by $\eta(0)=0$. Obviously, the code is at most $5$-weight. When $m=s+1$,
 note that $p^{2e-2}+p^{e+m+s-2}=p^{2e-3}(p-1)+2p^{e+m+s-2},$ thus the code is a four-weight linear code.

\begin{table}
\begin{center}
\caption{The weight distribution of $\mathcal{C}_D$ when $ m/s \equiv 0\,mod\,2 $}\label{2}
\begin{tabular}{ll}
\hline\noalign{\smallskip}
Weight  &  Multiplicity   \\
\noalign{\smallskip}
\hline\noalign{\smallskip}
$0$  &  1 \\
$p^{2e-2}+p^{e+m+s-2}$  &  $ p-1 $    \\
$(p-1)(p^{2e-3}+p^{e+m+s-3})$  &  $p^e(p^e-p^{1-2s})$     \\
$p^{2e-3}(p-1)$  &  $\frac{1}{2}(2p^{m-s-1}-3p^{m-s}+p^{m-s+1}+2p^{e-2s-1}-p^{e-2s}+p^{e-2s+1}-2)$     \\
$  p^{2e-3}(p-1)+2p^{e+m+s-2}$  &  $ \frac{1}{2}(2p^{m-s-1}-3p^{m-s}+p^{m-s+1}+2p^{e-2s-1}-3p^{e-2s}+p^{e-2s+1})$     \\
$ p^{2e-3}(p-1)+p^{e+m+s-2}$ & $ 1-p-2p^{m-s-1}+3p^{m-s}-p^{m-s+1}-2p^{e-2s-1}+2p^{e-2s} $     \\
\noalign{\smallskip}
\hline
\end{tabular}
\end{center}
\end{table}
\end{theorem}

The followings are some examples about our results verified by Magma.

\begin{example}\label{example1}
Let $(p,e,l)=(3, 2, 1)$, $\mathbb{F}_{3}^{*}=\langle 2 \rangle,$ then $m=1,$ $s=\gcd(e,l)=1,$ and $m/s \equiv 1\,mod\,2.$ By Theorem \ref{weight1}, the code $\mathcal{C}_D$ has parameters $[12,4,6]$ with weight enumerator $1+12z^{6}+54z^{8}+8z^{9}+6z^{12},$ and complete weight enumerator
\begin{displaymath}
\begin{aligned}
  & \omega_{0}^{12}+\omega_{1}^{12}+\omega_{2}^{12}+54(\omega_{0}\omega_{1}\omega_{2})^{4}
  +4\omega_{0}^{6}(\omega_{1}\omega_{2})^{3}+4\omega_{0}^{0}(\omega_{1}\omega_{2})^{6}\\
  &+4\omega_{0}^{6}\omega_{1}^{6}\omega_{2}^{0}+4\omega_{0}^{6}\omega_{1}^{0}\omega_{2}^{6}
  +4\omega_{0}^{3}\omega_{1}^{3}\omega_{2}^{6}+4\omega_{0}^{3}\omega_{1}^{6}\omega_{2}^{3}
\end{aligned}
\end{displaymath}
which confirmed the result by Magma. According to Griesmer bound, this code is almost optimal as the best linear code of length $12$ and dimension $4$ over $\mathbb{F}_{3}$ has minimum weight $7$. Furthermore, the code is optimal one with respect to the code table\cite{YZ}. In fact, we can get the same result by Magma when $l=3,5,7.$
\end{example}

\begin{example}\label{example2}
If $(p,e,l)=(3,4,1)$,  $\mathbb{F}_{3}^{*}=\langle 2 \rangle,$ then $m=2,$ $s=\gcd(e,l)=1,$ and $m/s \equiv 0\,mod\,2.$ By Theorem \ref{weight2}, the code $\mathcal{C}_D$ has parameters $[972,8,486]$ with weight enumerator
$1+12z^{486}+6534z^{648}+8z^{729}+6z^{972},$ and complete weight enumerator
\begin{displaymath}
\begin{aligned}
  &\omega_{0}^{972}+\omega_{1}^{972}+\omega_{2}^{972}+6534(\omega_{0}\omega_{1}\omega_{2})^{324}
  +4\omega_{0}^{486}(\omega_{1}\omega_{2})^{243}+4\omega_{0}^{0}(\omega_{1}\omega_{2})^{486}\\
  &+4\omega_{0}^{486}\omega_{1}^{486}\omega_{2}^{0}+4\omega_{0}^{486}\omega_{1}^{0}\omega_{2}^{486}
  +4\omega_{0}^{243}\omega_{1}^{243}\omega_{2}^{486}+4\omega_{0}^{243}\omega_{1}^{486}\omega_{2}^{243}
\end{aligned}
\end{displaymath}
which confirmed the result by Magma. In fact, we can get the same result by Magma when $l=3,5,7,9.$
\end{example}

\begin{example}\label{example3}
If $(p,e,l)=(3, 4, 2)$, $\mathbb{F}_{3}^{*}=\langle 2 \rangle,$ then $m=2,$ $s=\gcd(e,l)=2,$ and $m/s \equiv 1\,mod\,2.$ By Theorem \ref{weight1}, the code $\mathcal{C}_D$ has parameters $[810,8,486]$ with weight enumerator $1+110z^{486}+6318z^{540}+100z^{567}+30z^{648}+2z^{810},$ and complete weight enumerator
\begin{displaymath}
\begin{aligned}
  &\omega_{0}^{810}+\omega_{1}^{810}+\omega_{2}^{810}+6318(\omega_{0}\omega_{1}\omega_{2})^{270}
  +50\omega_{0}^{324}(\omega_{1}\omega_{2})^{243}+30\omega_{0}^{162}(\omega_{1}\omega_{2})^{324}\\
  &+30\omega_{0}^{324}\omega_{1}^{324}\omega_{2}^{162}+30\omega_{0}^{324}\omega_{1}^{162}\omega_{2}^{324}
  +50\omega_{0}^{243}\omega_{1}^{243}\omega_{2}^{324}+50\omega_{0}^{243}\omega_{1}^{324}\omega_{2}^{243}
\end{aligned}
\end{displaymath}
which confirmed the result by Magma. In fact, we can get the same result by Magma when $l=6,10.$
\end{example}

\section{Preliminaries and Auxiliary lemmas}\label{preliminaries}
In this section, we present some facts on exponential sums, that will be needed in calculating the complete weight enumerator of the codes defined in this article.

An additive character of $\mathbb{F}_q$ is a non-zero function $\chi$ from $\mathbb{F}_q$ to the set of complex numbers of absolute value $1$ such that
$\chi(x+y)=\chi(x)\chi(y)$ for any pair $(x,y) \in \mathbb{F}_q^2$. For each $u \in \mathbb{F}_q$, the function
$$\chi_u(v)=\zeta_{p}^{Tr(uv)},~v \in \mathbb{F}_q$$
denotes an additive character of $\mathbb{F}_q$, where $\zeta_{p}=e^{2\pi i/p}$ is a primitive $p$-th root of unity and $i=\sqrt{-1}$.  Since $\chi_0(v)=1$ for all $v \in \mathbb{F}_q$, which is the trivial additive character of $\mathbb{F}_q$. We call $\chi_1$ the canonical additive character of $\mathbb{F}_q$ and we have $\chi_u(x)=\chi_1(ux)$ for all $u\in\mathbb{F}_q$. The additive character satisfies the orthogonal property \cite{LN}, that is
\begin{eqnarray*}
\sum_{v \in \mathbb{F}_q} \chi_u(v)=\left\{
\begin{array}{ll}
q   & \mathrm{if}\,\ u=0,\\
0    & \mathrm{if}\,\ u\neq0. \\
\end{array}
\right.
\end{eqnarray*}

Let $h$ be a fixed primitive element of $\mathbb{F}_q$. For each $j=0, 1, \ldots, q-2,$ the function $\lambda_j(h^k)=e^{2\pi ijk/(q-1)}$ for $k=0, 1, \ldots, q-2$ defines a multiplicative character of $\mathbb{F}_q$, we extend these characters by setting $\lambda_j(0)=0$. Let $q$ be odd. For $j=(q-1)/2$ and $v \in \mathbb{F}_{q}^{*}$, we have
\begin{equation*}
  \lambda_{(q-1)/2}(v) =
  \begin{cases}
    1, & \text{if $v$ is the square of an element of $\mathbb{F}_{q}^{*}$,}  \\
    -1, & \text{otherwise,}
  \end{cases}
\end{equation*}
which is called the quadratic character of $\mathbb{F}_{q}$, and is denoted by $\eta'$ in the sequel.
We call $\eta'=\lambda_{(q-1)/2}$ and $\eta=\lambda_{(p-1)/2}$ are the quadratic characters over $\mathbb{F}_q$ and $\mathbb{F}_p$, respectively. The quadratic Gauss sums over $\mathbb{F}_q$ and $\mathbb{F}_p$ are defined respectively by
$$G'(\eta')=\sum\limits_{v \in \mathbb{F}_q}\eta'(v)\chi'_1(v) \quad \mathrm{and} \quad G(\eta)=\sum\limits_{v \in \mathbb{F}_p}\eta(v)\chi_1(v),$$
where $\eta$ and $\chi_1$ are the canonical multiplicative and additive characters of $\mathbb{F}_p$, respectively. Moreover, it is well known that $G'=(-1)^{e-1}\sqrt{p^{*}}^e$ and $G=\sqrt{p^{*}}$, where $p^{*}=\eta(-1)p.$

The following are some basic facts on exponential sums.

\begin{lemma}\label{quadratic sums}(\cite{LN}, Theorem 5.33)
If $f(x)=a_{2}x^{2}+a_{1}x+a_{0} \in \mathbb{F}_{q}[x],$ where $a_{2}\neq 0,$ then
$$\sum_{x \in \mathbb{F}_{q}}\zeta_{p}^{Tr(f(x))}=\zeta_{p}^{Tr(a_{0}-a_{1}^{2}(4a_{2})^{-1})}\eta'(a_{2})G'(\eta'),$$
where $\eta'$ is the quadratic character of $\mathbb{F}_{q}$.
\end{lemma}

\begin{lemma}\label{quadratic character}(\cite{LN}, Theorem 5.48)
With the notation above, we have
\begin{equation*}
  \sum_{x \in \mathbb{F}_{q}}\eta'(f(x)) =
  \begin{cases}
    -\eta'(a_{2}), & \text{if $a_{1}^{2}-4a_{0}a_{2}\neq 0,$}  \\
    (q-1)\eta'(a_{2}), & \text{if $a_{1}^{2}-4a_{0}a_{2}=0.$}
  \end{cases}
\end{equation*}
\end{lemma}

For $\alpha, \beta \in \mathbb{F}_{q}$ and any positive integer $l$, the Weil sums       $S(\alpha, \beta)$ is defined by
$$S(\alpha, \beta)= \sum_{x \in \mathbb{F}_{q}}\zeta_{p}^{Tr(\alpha x^{p^{l}+1}+\beta x)}.$$

We will show some results of $S(\alpha, \beta)$ for $\alpha\neq 0$ and $q$ odd.

\begin{lemma}\label{weil sums1}(\cite{S}, Theorem 2)
Let $s=(l,e)$ and $e/s$ be even with $e=2m$. Then
\begin{equation*}
  S(\alpha, 0) =
  \begin{cases}
   (-1)^{m/s}p^m , & \text{if $\alpha^{(q-1)/(p^s+1)} \neq (-1)^{m/s},$}  \\
    (-1)^{m/s+1}p^{m+s}, & \text{if $\alpha^{(q-1)/(p^s+1)}=(-1)^{m/s}.$}
  \end{cases}
\end{equation*}
\end{lemma}

\begin{lemma}\label{weil sums2}(\cite{C}, Theorem 4.7)
Let $\beta \neq 0$ and $e/s$ be even with $e=2m.$ Then $S(\alpha, \beta)=0$ unless the equation $\alpha^{p^{l}}X^{p^{2l}}+\alpha X=-\beta^{p^{l}}$ is solvable. There are two possibilities.
\begin{enumerate}
  \item If $\alpha^{(q-1)/(p^s+1)} \neq (-1)^{m/s},$ then for any choice of $\beta \in \mathbb{F}_q,$ the equation has a unique solution $x_{0}$ and $$S(\alpha, \beta)=(-1)^{m/s}p^{m}\zeta_{p}^{Tr(-\alpha x_{0}^{p^{l}+1})}$$
  \item If $\alpha^{(q-1)/(p^s+1)}=(-1)^{m/s}$ and if the equation is solvable with some solution $x_{0}$, then
  $$S(\alpha, \beta)=(-1)^{m/s+1}p^{m+s}\zeta_{p}^{Tr(-\alpha x_{0}^{p^{l}+1})}$$
\end{enumerate}
\end{lemma}

\begin{lemma}\label{weil sums3}(\cite{S}, Theorem 4.1)
For $e=2m$, the equation $\alpha^{p^{l}} X^{p^{2l}}+\alpha X=0$ is solvable for $X \in \mathbb{F}_{q}^{*}$ if and only if $e/s$ is even and $\alpha^{(q-1)/(p^{s}+1)}=(-1)^{m/s}$. In such cases, there are $p^{2s}-1$ non-zero solutions.
\end{lemma}

There is the fact that $\alpha^{p^{l}}X^{p^{2l}}+\alpha X$ is a permutation polynomial over $\mathbb{F}_{q}$ with $q=p^e$ if and only if $e/s$ is odd or $e/s$ is even with $e=2m$ and $\alpha^{(q-1)/(p^s+1)} \neq (-1)^{m/s}$.

\begin{lemma}\label{weil sums4}(\cite{XS,QFKD})
Let $f(X)=X^{p^{2l}}+X$ and $$S=\{\beta \in \mathbb{F}_{q}: f(X)=-\beta^{p^{l}} is\;solvable\;in\;\mathbb{F}_{q}\}.$$
If $m/s \equiv 0\,mod\,2$, then $|S|=p^{e-2s}$.
\end{lemma}
\textbf{proof:} Take into account that both $e/s$ and $m/s$ are even, we can deduce that $\alpha^{(q-1)/(p^s+1)}=(-1)^{m/s}$.
Then $f(X)=0$ has $p^{2s}$ solutions in $\mathbb{F}_{q}$ from Lemma \ref{weil sums3}. So does the equation $f(X)=-\beta^{p^l}$ with $\beta \in S$. For $\beta_{1}, \beta_{2} \in \mathbb{F}_{q}$ and $\beta_{1} \neq \beta_{2}$, there are no common solutions for equations $f(X)=-\beta_{1}^{p^l}$ and $f(X)=-\beta_{2}^{p^l}$. In addition, for each $\alpha \in \mathbb{F}_{q}$, it is known that $f(\alpha)$ is in $\mathbb{F}_{q}$, there must be some $\beta \in \mathbb{F}_{q}$ such that $f(\alpha)=-\beta^{p^l}$. Since by use of the equation $|S|\cdot p^{2s} =p^e$, we can get the desired conclusion. \hfill$\square$

\section{The proofs of the main results}\label{proofs}

The following Lemmas \ref{N(u,v)}-\ref{T2} are essential to determine the lengths, complete weight enumerators                                                                                                                                                                                                                                                                                                                                                                                                                                                                                                                                                       and weight distributions of $\mathcal{C}_D$.

\begin{lemma}\label{N(u,v)}
The length of the code $\mathcal{C}_D$ is
\begin{equation*}
\begin{aligned}
  n&=|\{ (x_1,x_2) \in \mathbb{F}_q^2:Tr(x_2) = 1,Tr(x_1^{p^l + 1}) = 1\}|\\&=
  \begin{cases}
    p^{2e-2}+p^{e+m-2}, & \text{if $m/s\equiv 1\,mod\,2$,}  \\
    p^{2e-2}+p^{e+m+s-2}, & \text{if $m/s\equiv 0\,mod\,2$. }
  \end{cases}
  \end{aligned}
\end{equation*}
\end{lemma}
\textbf{Proof:} Since
\begin{eqnarray*}
n&=&|\{ (x_1,x_2) \in \mathbb{F}_q^2:Tr(x_2) = 1,Tr(x_1^{p^l + 1}) = 1\}|\\
&=& \sum_{x_1, x_2 \in \mathbb{F}_q}(\frac{1}{p} \sum_{y_1 \in \mathbb{F}_p}\zeta ^{y_1(Tr(x_1^{p^l+1}) - 1)} )(\frac{1}{p}\sum_{y_2 \in \mathbb{F}_p} \zeta^{y_2(Tr(x_2) - 1)}) \\
&=& p^{2e - 2} + \frac{1}{p^2}(A_1 + A_2 + A_3 ), \\
\end{eqnarray*}
where
\begin{eqnarray*}
A_1 &=& \sum_{x_1,x_2 \in \mathbb{F}_q}\sum_{y_2 \in \mathbb{F}_p^*}\zeta_{p}^{y_2Tr(x_2)-y_2} = \sum_{y _2 \in \mathbb{F}_p^*}\zeta_{p}^{-y_2} \sum_{x_1,x_2 \in \mathbb{F}_q}\zeta_{p}^{y_2Tr(x_2)} = 0,\\
A_2 &=& \sum_{x_1,x_2\in \mathbb{F}_q} \sum_{y_1 \in \mathbb{F}_p^*} \zeta_{p}^{y_1Tr(x_1^{p^l + 1}) - y_1}=\sum_{y_1 \in\mathbb{F}_p^*} \zeta_{p}^{- y_1} \sum_{x_1 \in \mathbb{F}_q} \zeta_{p}^{y_1Tr(x_1^{p^l + 1})}\sum_{x_{2} \in \mathbb{F}_{q}}1\\&=&p^e\sum_{y_1 \in\mathbb{F}_p^*} \zeta_{p}^{- y_1} \sum_{x_1 \in \mathbb{F}_q} \zeta_{p}^{y_1Tr(x_1^{p^l + 1})} = p^e\sum_{y_{1} \in \mathbb{F}_p^*} \zeta_{p}^{-y_{1}} \cdot S(y_{1},0)\\
&=&
\begin{cases}
  p^{e+m}, & \mbox{if } m/s \equiv 1\,mod\,2; \\
  p^{e+m+s}, & \mbox{if } m/s \equiv 0\,mod\,2.
\end{cases}\\
A_3&=& \sum_{x_1,x_2\in \mathbb{F}_q}\sum_{y_1 \in \mathbb{F}_p^*}\zeta_{p}^{y_1Tr(x_1^{p^l + 1}) - y_1}\sum_{y_2 \in \mathbb{F}_p^*} \zeta_{p}^{y_2Tr(x_2) - y_2}\\
&=&\sum_{y_1 \in \mathbb{F}_p^*}\zeta_{p}^{ -y_1} \sum_{y_2 \in \mathbb{F}_p^*} \zeta_{p}^{ - y_2}\sum_{x_1\in \mathbb{F}_q}\zeta_{p}^{y_1Tr(x_1^{p^l + 1})}\sum_{x_2\in \mathbb{F}_q}\zeta_{p}^{y_2Tr(x_2)}\\
&=&0.
\end{eqnarray*}
Then
\begin{equation*}
  n=
  \begin{cases}
    p^{2e-2}+p^{e+m-2}, & \mbox{if } m/s \equiv 1\,mod\,2;\\
    p^{2e-2}+p^{e+m+s-2}, & \mbox{if } m/s \equiv 0\,mod\,2.
  \end{cases}
\end{equation*}

The results of $A_1,$ $A_2$ and $A_3$ are all due to orthogonal property of the additive character. Furthermore, the result of $A_2$ is also based on the Lemma \ref{weil sums1}. From all the discussions above, we complete the proof of the lemma.\hfill$\square$

For any $a, b\in \mathbb{F}_q$ and $\mathbf{c}(a, b)\in \mathcal{C}_D$, to determine the complete weight enumerator of $\mathcal{C}_{D}$, let $N_{\rho}(a,b)$ denote the number of components $Tr(ax_1+bx_2)$ of codeword $\mathbf{c}(a, b)$ that equal to $\rho$, where $\rho \in \mathbb{F}_{p}$, $a,b \in \mathbb{F}_{q}$.
We define
$$N_{\rho}(a,b)=|\{ (x_1, x_2) \in \mathbb{F}_q^2:Tr(x_1^{p^l + 1}) = 1, Tr(x_2) = 1,Tr(ax_1+bx_2) = \rho, x_1, x_2\in \mathbb{F}_q \}.$$
Then it is not difficult to obtain the Hamming weight of $\mathbf{c}(a, b)$, that is
\begin{eqnarray}
wt(\mathbf{c}(a,b)) = \sum_{\rho \in \mathbb{F}_{p}^{*}}N_{\rho}(a,b) = n - N_{0}(a,b).
\end{eqnarray}
Here we only  need to consider $\rho \in \mathbb{F}_{p}^{*}$ in the sequel.

By the definition of $N_{\rho}(a,b)$, we have
\begin{eqnarray}\label{5}
N_{\rho}(a,b) &=& \sum_{x_1, x_2 \in \mathbb{F}_q}(\frac{1}{p} \sum_{z_1 \in \mathbb{F}_p} \zeta_{p}^{z_1Tr(x_1^{p^l + 1}) - z_1})(\frac{1}{p}\sum_{z_2 \in \mathbb{F}_p} \zeta_{p}^{z_2Tr(x_2)-z_2})(\frac{1}{p}\sum_{ z_3 \in \mathbb{F}_p} \zeta_{p}^{z_3 (Tr(ax_1+bx_2)-\rho)} ) \nonumber\\
&=& \frac{n}{p} +\varphi_1 + \varphi_2+ \varphi_3+ \varphi_4,
\end{eqnarray}
where
\begin{eqnarray*}
\varphi _1 &=& p^{-3}\sum_{x_1,x_2 \in \mathbb{F}_q} \sum_{z_3  \in \mathbb{F}_p^*}\zeta_{p}^{z_3 (Tr(ax_1+bx_2)-\rho)}\\
&=& p^{-3}\sum_{z_3  \in \mathbb{F}_p^*}\zeta_{p}^{-\rho z_3} \sum_{x_1 \in \mathbb{F}_q}\zeta_{p}^{ Tr(a z_3 x_1)}\sum_{x_2 \in \mathbb{F}_q}\zeta_{p}^{ Tr(b z_3 x_2)}\\
&=&
\begin{cases}
  -p^{2e-3}, & \mbox{if } a=b=0; \\
  0, & \mbox{otherwise}.
\end{cases}
\end{eqnarray*}


\begin{eqnarray*}
\varphi _2& = & p^{-3}\sum_{x_1,x_2 \in \mathbb{F}_q} \sum_{z_2 \in \mathbb{F}_p^*} \zeta_{p} ^{z_2(Tr(x_2) - 1)} \sum_{z_3  \in \mathbb{F}_p^*} \zeta_{p}^{z_3 (Tr(ax_1+bx_2)-\rho)}\\
&=& p^{-3} \sum_{z_2 \in \mathbb{F}_p^*} \zeta_{p} ^{-z_2} \sum_{z_3  \in \mathbb{F}_p^*}\zeta_{p}^{-\rho z_3}\sum_{x_2\in \mathbb{F}_q}\zeta_{p}^{ Tr((z_2+bz_3)x_2)}\sum_{x_1\in \mathbb{F}_q}\zeta_{p}^{ Tr(az_3x_1)}\\
&=& \left\{
\begin{array}{ll}
  p^{e-3}\sum\limits_{z_2 \in \mathbb{F}_p^*} \zeta_{p} ^{-z_2} \sum\limits_{z_3  \in \mathbb{F}_p^*}\zeta_{p}^{-\rho z_3}\sum\limits_{x_2\in \mathbb{F}_q}\zeta_{p}^{ (Tr(z_2+bz_3)x_2)}, & \mbox{if } a=0; \\
  0, & \mbox{if } a \neq 0.
\end{array}
\right.\\
&=& \left\{
\begin{array}{ll}
p^{e-3}\sum\limits_{z_2 \in \mathbb{F}_p^*} \zeta_{p} ^{-z_2} (\zeta_{p}^{-\rho(-b^{-1}z_{2})}\sum\limits_{x_{2} \in \mathbb{F}_q}\zeta_{p}^{0}+\sum\limits_{z_3 \neq -b^{-1}z_{2}}\zeta_{p}^{-\rho z_3}\sum\limits_{x_2\in \mathbb{F}_q}\zeta_{p}^{ (Tr(z_2+bz_3)x_2)}), \\ \qquad \qquad \qquad \qquad \qquad \qquad \qquad \qquad \qquad \qquad \mbox{if } a=0\:and\:b \in \mathbb{F}_{p}^*; \\
 0, \qquad \qquad \qquad \qquad \qquad \qquad \qquad \qquad \qquad \qquad \mbox{otherwise } .
\end{array}
\right.\\
&=& \left\{
\begin{array}{ll}
p^{2e-3}\sum\limits_{z_2 \in \mathbb{F}_p^*} \zeta_{p}^{(\rho b^{-1}-1)z_2}, & \mbox{if } a=0\:and\:b \in \mathbb{F}_{p}^*; \\
 0, & \mbox{otherwise } .
\end{array}
\right.\\
&=& \left\{
\begin{array}{ll}
p^{2e-3}(p-1), & \mbox{if } a=0\:and\:\rho=b ; \\
-p^{2e-3}, & \mbox{if } a=0\:and\:\rho \neq b ; \\
 0, & \mbox{otherwise } .
\end{array}
\right.\\
\end{eqnarray*}

The results of $\varphi_1$ and $\varphi_2$ are derived from the orthogonal property of the additive character. And in order to determine the values of $\varphi_3$, $\varphi_{4}$, we have the following lemmas.

\begin{eqnarray*}
\varphi _3 &=& p^{-3}\sum_{x_1,x_2 \in \mathbb{F}_q} \sum_{z_1 \in \mathbb{F}_p^*} \zeta_{p} ^{z_1(Tr(x_1^{p^l + 1})-1)} \sum_{z_3  \in \mathbb{F}_p^*} \zeta_{p} ^{z_3 (Tr(ax_1+bx_2)-\rho)}\\
&=&p^{-3}\sum_{z_1 \in \mathbb{F}_p^*} \zeta_{p} ^{-z_1}\sum_{z_3 \in \mathbb{F}_p^*}\zeta_{p} ^{-\rho z_3}\sum_{x_1\in \mathbb{F}_q}\zeta_{p} ^{Tr(z_1x_1^{p^l+1}+az_3x_1)}\sum_{x_2\in \mathbb{F}_q}\zeta_{p} ^{Tr(bz_3x_2)}   \\
&=& \left\{
\begin{array}{ll}
p^{e-3}\sum\limits_{z_1 \in \mathbb{F}_p^*} \zeta_{p} ^{-z_1}\sum\limits_{z_3  \in \mathbb{F}_p^*} \zeta_{p}^{-\rho z_3}\sum\limits_{x_1\in \mathbb{F}_q}\zeta_{p} ^{Tr(z_1x_1^{p^l+1}+az_3x_1)}, & \mbox{if } b = 0, \\
0, & \mbox{if }  b \ne 0,
\end{array}
\right.\\
\end{eqnarray*}

Next, we will deduce all the other remaining solutions of $\varphi_3$ through three lemmas.

\begin{lemma}\label{CWE1}
With the notations above, if $b=0$ and $a=0$, we have
\begin{equation*}
\varphi_{3}=
\begin{cases}
-p^{e+m-3}, & \mbox{when } m/s \equiv 1 \, mod \, 2;\\
-p^{e+m+s-3}, & \mbox{when } m/s \equiv 0 \, mod \, 2.
\end{cases}
\end{equation*}
\end{lemma}
\textbf{Proof:} If $a=0,$ then $\sum\limits_{x_1 \in \mathbb{F}_{q}}\zeta_{p}^{Tr(z_1 x_1^{p^l+1}+az_3x_1)}=\sum\limits_{x_1 \in \mathbb{F}_q}\zeta_{p}^{Tr(z_1x_1^{p^l+1})}=S(z_1, 0)$.
Since $e/s$ is even, $e=2m$, there exist some $t_0$, such that $e=s \cdot 2t_0$. As $z_1 \in \mathbb{F}_{p}^{*}$, then
\begin{displaymath}
  \begin{aligned}
  z_1^{\frac{q-1}{p^s+1}}&=z_1^{\frac{p^e-1}{p^s+1}}=z_1^{\frac{(p^{2s})^{t_0}-1}{p^s+1}}
  =z_1^{\frac{(p^{2s}-1)(p^{2s(t_0-1)}+p^{2s(t_0-2)}+\ldots+1)}{p^s+1}}=z_1^{(p^s-1)(p^{2s(t_0-1)}+p^{2s(t_0-2)}+\ldots+1)}\\
  &=z_1^{(p-1)(p^{s-1}+p^{s-2}+\ldots+1)(p^{2s(t_0-1)}+p^{2s(t_0-2)}+\ldots+1)}=1.
  \end{aligned}
\end{displaymath}
\begin{enumerate}
  \item If $m/s \equiv 1\,mod\,2$, $z_1^{\frac{q-1}{p^s+1}} \neq (-1)^{m/s}$. By means of Lemma \ref{weil sums1}, $S(z_1, 0)=-p^{m}$. Then
      \begin{displaymath}
        \begin{aligned}
        \varphi_3&=p^{e-3}\sum\limits_{z_1 \in \mathbb{F}_{p}^*}\zeta_{p}^{-z_1}\sum\limits_{z_3 \in \mathbb{F}_{p}^*}\zeta_{p}^{-\rho z_3}S(z_1,0)\\
        &=p^{e-3} \cdot (-1) \cdot (-1) \cdot (-p^{m})=-p^{e+m-3},
        \end{aligned}
      \end{displaymath}
  \item If $m/s \equiv 0\,mod\,2$, $z_1^{\frac{q-1}{p^s+1}} = (-1)^{m/s}$. By means of Lemma \ref{weil sums1}, $S(z_1, 0)=-p^{m+s}$. Then
      \begin{displaymath}
        \begin{aligned}
        \varphi_3&=p^{e-3}\sum\limits_{z_1 \in \mathbb{F}_{p}^*}\zeta_{p}^{-z_1}\sum\limits_{z_3 \in \mathbb{F}_{p}^*}\zeta_{p}^{-\rho z_3}S(z_1,0)\\
        &=p^{e-3} \cdot (-1) \cdot (-1) \cdot (-p^{m+s})=-p^{e+m+s-3},
        \end{aligned}
      \end{displaymath}
\end{enumerate}
which lead to the desired conclusion. \hfill$\square$
\begin{lemma}\label{CWE2}
With the notations above, if $b=0$, $a \neq 0$, $m/s \equiv 1 \, mod \, 2$, then the equation $X^{p^{2l}}+X=-a^{p^l}$
has a unique solution $\gamma$ in $\mathbb{F}_{q}$, and
\begin{equation*}
\varphi_{3}=
\begin{cases}
-p^{e+m-3}-p^{e+m-2}\eta(\rho^2-4Tr(\gamma^{p^l+1})), & \mbox{when } Tr(\gamma^{p^l+1}) \neq 0 ;\\
-p^{e+m-3}, & \mbox{when } Tr(\gamma^{p^l+1}) = 0 .
\end{cases}
\end{equation*}
\end{lemma}
\textbf{Proof:} If $m/s \equiv 1 \, mod \, 2$, satisfy that $\alpha^{\frac{q-1}{p^s+1}} \neq (-1)^{m/s}$, since $\alpha=1$. By use of Lemma \ref{weil sums3}, it is clearly that $X^{p^{2l}}+X$ is a permutation polynomial over $\mathbb{F}_q$ and $X^{p^{2l}}+X=-a^{p^l}$ has a unique solution $\gamma$ in $\mathbb{F}_{q}$. And $z_3z_1^{-1}\gamma$ is the unique solution of the equation $(z_1X)^{p^{2l}}+z_1X=-(az_3)^{p^{l}}$ for any $z_1,z_3 \in \mathbb{F}_{p}^*$. By Lemma \ref{weil sums2},
\begin{displaymath}
\begin{aligned}
  S(z_1, az_3)&=(-1)^{m/s}p^m \zeta_{p}^{Tr(-z_1(z_3z_1^{-1}\gamma)^{p^l+1})}\\
              &=(-1)^{m/s}p^m \zeta_{p}^{-\frac{z_3^2}{z_1}Tr(\gamma^{p^l+1})}\\&=
              \begin{cases}
                -p^m \zeta_{p}^{-\frac{z_3^2}{z_1}Tr(\gamma^{p^l+1})}, & \mbox{if } Tr(\gamma^{p^l+1}) \neq 0 \\
                -p^m, & \mbox{if } Tr(\gamma^{p^l+1})=0.
              \end{cases}
\end{aligned}
\end{displaymath}
\begin{enumerate}
  \item If $Tr(\gamma^{p^l+1}) \neq 0$,
    \begin{displaymath}
      \begin{aligned}
      \varphi_3 &=p^{e-3}\sum\limits_{z_1 \in \mathbb{F}_{p}^{*}}\zeta_{p}^{-z_1}\sum\limits_{z_3 \in \mathbb{F}_{p}^{*}}\zeta_{p}^{-\rho z_3}S(z_1, az_3)\\
                &=-p^{e+m-3}\sum\limits_{z_1 \in \mathbb{F}_{p}^{*}}\zeta_{p}^{-z_1}\sum\limits_{z_3 \in \mathbb{F}_{p}^{*}}\zeta_{p}^{-\frac{z_3^2}{z_1}Tr(\gamma^{p^l+1})-\rho z_3}\\
                &=-p^{e+m-3}\sum\limits_{z_1 \in \mathbb{F}_{p}^{*}}\zeta_{p}^{-z_1}(\sum\limits_{z_3 \in \mathbb{F}_{p}}\zeta_{p}^{-\frac{z_3^2}{z_1}Tr(\gamma^{p^l+1})-\rho z_3}-1)\\
                &=-p^{e+m-3}\sum\limits_{z_1 \in \mathbb{F}_{p}^{*}}\zeta_{p}^{-z_1}\sum\limits_{z_3 \in \mathbb{F}_{p}}\zeta_{p}^{-\frac{z_3^2}{z_1}Tr(\gamma^{p^l+1})-\rho z_3}+p^{e+m-3}\sum\limits_{z_1 \in \mathbb{F}_{p}^{*}}\zeta_{p}^{-z_1}\\
                &=-p^{e+m-3}\sum\limits_{z_1 \in \mathbb{F}_{p}^{*}}\zeta_{p}^{-z_1}\zeta_{p}^{-\frac{(-\rho)^2}{4(-\frac{1}{z_1}Tr(\gamma^{p^l+1}))}}
                \eta(-\frac{1}{z_1}Tr(\gamma^{p^l+1}))G(\eta)-p^{e+m-3}\\
                &=-p^{e+m-3}\sum\limits_{z_1 \in \mathbb{F}_{p}^*}\zeta_{p}^{\frac{\rho^2-4Tr(\gamma^{p^l+1})}{4Tr(\gamma^{p^l+1})}z_1}
                \eta(-\frac{1}{z_1}Tr(\gamma^{p^l+1}))G(\eta)-p^{e+m-3}\\
                &=-p^{e+m-3}(G(\eta))^2\eta(-1)\eta(\rho^2-4Tr(\gamma^{p^l+1}))-p^{e+m-3}\\
                &=-p^{e+m-2}\eta(\rho^2-4Tr(\gamma^{p^l+1}))-p^{e+m-3},
      \end{aligned}
    \end{displaymath}
    where the fifth identity is due to Lemma \ref{quadratic sums}, while the last two formulas are based on the definition and property of quadratic Gauss sums over $\mathbb{F}_{p}$.
  \item If $Tr(\gamma^{p^l+1}) = 0$,
  \begin{displaymath}
    \begin{aligned}
    \varphi_3 &=p^{e-3}\sum\limits_{z_1 \in \mathbb{F}_{p}^{*}}\zeta_{p}^{-z_1}\sum\limits_{z_3 \in \mathbb{F}_{p}^{*}}\zeta_{p}^{-\rho z_3}S(z_1, az_3)\\
              &=p^{e-3}\sum\limits_{z_1 \in \mathbb{F}_{p}^{*}}\zeta_{p}^{-z_1}\sum\limits_{z_3 \in \mathbb{F}_{p}^{*}}\zeta_{p}^{-\rho z_3} \cdot (-p^m)\\
              &=-p^{e+m-3}\sum\limits_{z_1 \in \mathbb{F}_{p}^{*}}\zeta_{p}^{-z_1}\sum\limits_{z_3 \in \mathbb{F}_{p}^{*}}\zeta_{p}^{-\rho z_3}\\
              &=-p^{e+m-3},
    \end{aligned}
  \end{displaymath}
  where the result is derived from the orthogonal property of the additive character.
\end{enumerate}

Then we can get the desired conclusion.\hfill$\square$

\begin{lemma}\label{CWE3}
With the notations above, let $b=0$, $a \neq 0$, $m/s \equiv 0 \, mod \, 2$. If the equation $X^{p^{2l}}+X=-a^{p^l}$
has no solution in $\mathbb{F}_{q}$, then $\varphi_3=0$. Suppose that $X^{p^{2l}}+X=-a^{p^l}$
has a solution $\gamma$ in $\mathbb{F}_{q}$, we have
\begin{equation*}
\varphi_{3}=
\begin{cases}
-p^{e+m+s-3}-p^{e+m+s-2}\eta(\rho^2-4Tr(\gamma^{p^l+1})), & \mbox{when } Tr(\gamma^{p^l+1}) \neq 0 ;\\
-p^{e+m+s-3}, & \mbox{when } Tr(\gamma^{p^l+1}) = 0 .
\end{cases}
\end{equation*}
\end{lemma}
\textbf{Proof:} If $m/s \equiv 0 \, mod \, 2$, satisfy that $\alpha^{\frac{q-1}{p^s+1}} = (-1)^{m/s}$, since $\alpha=1$. By Lemma \ref{weil sums2}, $S(z_1, az_3)=0$ unless the equation $X^{p^{2l}}+X=-a^{p^l}$ is solvable. From Lemma \ref{weil sums3}, $f(X)=X^{p^{2l}}+X$ is not permutation polynomial over $\mathbb{F}_q$. If the equation $X^{p^{2l}}+X=-a^{p^l}$
has a solution $\gamma$ in $\mathbb{F}_{q}$, with Lemma \ref{weil sums2},
\begin{displaymath}
\begin{aligned}
  S(z_1, az_3)&=(-1)^{m/s+1}p^{m+s} \zeta_{p}^{Tr(-z_1(z_3z_1^{-1}\gamma)^{p^l+1})}\\
              &=(-1)^{m/s+1}p^{m+s} \zeta_{p}^{-\frac{z_3^2}{z_1}Tr(\gamma^{p^l+1})}\\&=
              \begin{cases}
                -p^{m+s} \zeta_{p}^{-\frac{z_3^2}{z_1}Tr(\gamma^{p^l+1})}, & \mbox{if } Tr(\gamma^{p^l+1}) \neq 0 \\
                -p^{m+s}, & \mbox{if } Tr(\gamma^{p^l+1})=0.
              \end{cases}
\end{aligned}
\end{displaymath}
The following proof process is similar to that of Lemma \ref{CWE2}, and will not be included here.\hfill$\square$

\begin{eqnarray*}
\varphi _4& =& p^{-3}\sum_{x_1,x_2 \in \mathbb{F}_q} \sum_{z_1 \in \mathbb{F}_p^*} \zeta_{p} ^{z_1(Tr(x_1^{p^l+1} - 1)} \sum_{z_2 \in \mathbb{F}_p^*} \zeta_{p}^{z_2(Tr(x_2)-1)} \sum_{z_3  \in \mathbb{F}_p^*} \zeta_{p}^{z_3 (Tr(ax_1+bx_2)-\rho)}\\
&=&p^{-3}\sum_{z_1 \in \mathbb{F}_p^*} \zeta_{p} ^{-z_1}\sum_{z_2 \in \mathbb{F}_p^*} \zeta_{p} ^{-z_2}\sum_{z_3 \in \mathbb{F}_p^*}\zeta_{p}^{-\rho z_3}\sum_{x_1 \in \mathbb{F}_q} \zeta_{p}^{Tr(z_1x_1^{p^l+1}+az_3x_1)} \sum_{x_2 \in \mathbb{F}_q} \zeta_{p}^{Tr((z_2+bz_3)x_2)}\\
\end{eqnarray*}
\begin{enumerate}
  \item If $b \in \mathbb{F}_{q}/\mathbb{F}_{p}^*$, since $z_2, z_3 \in \mathbb{F}_{p}^*$, we have $z_2+bz_3 \neq 0$, then $\varphi_4=0$;
  \item If $b \in \mathbb{F}_{p}^*$, $z_2, z_3 \in \mathbb{F}_{p}^*$, for some $b$ such that $z_2+bz_3=0$, then
  \begin{displaymath}
    \begin{aligned}
    \varphi_4&=p^{-3}\sum_{z_1 \in \mathbb{F}_{p}^*}\zeta_{p}^{-z_1}\sum_{z_2 \in \mathbb{F}_{p}^*}\zeta_{p}^{-z_2}(\zeta_{p}^{-\rho(-b^{-1}z_2)}\sum_{x_1 \in \mathbb{F}_{q}}\zeta_{p}^{Tr(z_1x_1^{p^l+1}+a(-b^{-1}z_2)x_1)}\sum_{x_2 \in \mathbb{F}_q}\zeta_{p}^0\\
    &+\sum_{z_3 \neq -b^{-1}z_2}\zeta_{p}^{-\rho z_{3}}\sum_{x_1 \in \mathbb{F}_q}\zeta_{p}^{Tr(z_1x_1^{p^l+1}+az_3x_1)}\sum_{x_2 \in \mathbb{F}_q}\zeta_{p}^{Tr((z_2+bz_3)x_2)})\\
    &=p^{e-3}\sum_{z_1 \in \mathbb{F}_p^*}\zeta_{p}^{-z_1}\sum_{z_2 \in \mathbb{F}_p^*}\zeta_{p}^{(\rho b^{-1}-1)z_2}\sum_{x_1 \in \mathbb{F}_q}\zeta_{p}^{Tr(z_1x_1^{p^l+1}+(-ab^{-1}z_2)x_1)}\\
    &=p^{e-3}\sum_{z_1 \in \mathbb{F}_p^*}\zeta_{p}^{-z_1}\sum_{z_2 \in \mathbb{F}_p^*}\zeta_{p}^{(\rho b^{-1}-1)z_2}S(z_1, -ab^{-1}z_2),
    \end{aligned}
  \end{displaymath}
  where the second identity is from the orthogonal property of the additive character, and the last identity is based on the definition of Weil sums.
\end{enumerate}

In the sequel, we will give the rest values of $\varphi_4$ through the other three lemmas.

\begin{lemma}\label{CWE4}
With the notations above, if $b \in \mathbb{F}_p^*$ and $a=0$, we have
\begin{equation*}
\varphi_{4}=
\begin{cases}
-p^{e+m-3}, & \mbox{when } m/s \equiv 1 \, mod \, 2 \:and \: \rho \neq b;\\
p^{e+m-3}(p-1), & \mbox{when } m/s \equiv 1 \, mod \, 2 \:and \: \rho = b;\\
-p^{e+m+s-3}, & \mbox{when } m/s \equiv 0 \, mod \, 2 \:and \: \rho \neq b;\\
p^{e+m+s-3}(p-1), & \mbox{when } m/s \equiv 0 \, mod \, 2 \:and \: \rho = b.
\end{cases}
\end{equation*}
\end{lemma}
\textbf{Proof:} In this case, with Lemma \ref{weil sums1} and the orthogonal property of the additive character, the proof of this lemma is analogous to that in Lemma \ref{CWE1}. The details are omitted.\hfill$\square$

\begin{lemma}\label{CWE5}
With the notations above, if $b \in \mathbb{F}_{p}^*$, $a \neq 0$, $m/s \equiv 1 \, mod \, 2$, then the equation $X^{p^{2l}}+X=-a^{p^l}$
has a unique solution $\gamma$ in $\mathbb{F}_{q}$, and
\begin{equation*}
\varphi_{4}=
\begin{cases}
p^{e+m-3}(p-1), & \mbox{when } Tr(\gamma^{p^l+1}) = 0\:and\:\rho = b;\\
-p^{e+m-3}, & \mbox{when } Tr(\gamma^{p^l+1}) = 0\:and\:\rho \neq b;\\
-p^{e+m-3}-p^{e+m-2}\eta((\rho-b)^2-4Tr(\gamma^{p^l+1})), & \mbox{when } Tr(\gamma^{p^l+1}) \neq 0.
\end{cases}
\end{equation*}
\end{lemma}
\textbf{Proof:} With Lemma \ref{weil sums2}, by means of the similar proof process of Lemma \ref{CWE2}, we can deduce the desired result and will not be included here.\hfill$\square$

\begin{lemma}\label{CWE6}
With the notations above, let $b \in \mathbb{F}_p^*$, $a \neq 0$, $m/s \equiv 0 \, mod \, 2$. If the equation $X^{p^{2l}}+X=-a^{p^l}$
has no solution in $\mathbb{F}_{q}$, then $\varphi_4=0$. Suppose that $X^{p^{2l}}+X=-a^{p^l}$
has a solution $\gamma$ in $\mathbb{F}_{q}$, we have
\begin{equation*}
\varphi_{4}=
\begin{cases}
p^{e+m+s-3}(p-1), & \mbox{when } Tr(\gamma^{p^l+1}) = 0\:and\:\rho = b;\\
-p^{e+m+s-3}, & \mbox{when } Tr(\gamma^{p^l+1}) = 0\:and\:\rho \neq b;\\
-p^{e+m+s-3}-p^{e+m+s-2}\eta((\rho-b)^2-4Tr(\gamma^{p^l+1})), & \mbox{when } Tr(\gamma^{p^l+1}) \neq 0.
\end{cases}
\end{equation*}
\end{lemma}
\textbf{Proof:} By Lemma \ref{weil sums2}, we can also give the proof similar to Lemma \ref{CWE3} and we will not demonstrate it in detail here.\hfill$\square$

Based on the discussion above, we can get the values of $N_{\rho}(a,b)$.
\begin{lemma}\label{T1}
With the notations above and $m/s \equiv 1\,mod\,2$, we have
\begin{enumerate}
  \item If $a = 0, b = 0$, then $N_{\rho}(a,b)=0$.
  \item If $a = 0, b \neq 0$,
  \begin{enumerate}
    \item[(1)] $b \in \mathbb{F}_{p}^*$, then
    \begin{equation*}
      N_{\rho}(a,b)=
      \begin{cases}
      p^{2e-2}+p^{e+m-2}, & \mbox{when } \rho = b; \\
      0, & \mbox{when } \rho \neq b.
    \end{cases}
    \end{equation*}
    \item[(2)] $b \in \mathbb{F}_{q}^*/\mathbb{F}_{p}^*$, then
    $N_{\rho}(a,b)=p^{2e-3}+p^{e+m-3}$.
  \end{enumerate}
  \item If $a \neq 0, b = 0$, then
  \begin{equation*}
      N_{\rho}(a,b)=
      \begin{cases}
      p^{2e-3}, & \mbox{when } Tr(\gamma^{p^l+1}) = 0; \\
      p^{2e-3}-p^{e+m-2}\eta(\rho^2-4Tr(\gamma^{p^l+1})), & \mbox{when } Tr(\gamma^{p^l+1}) \neq 0.
    \end{cases}
    \end{equation*}
  \item If $a \neq 0, b \neq 0$,
  \begin{enumerate}
    \item[(1)]  $b \in \mathbb{F}_{p}^*$,
    \begin{enumerate}
      \item[(i)] $Tr(\gamma^{p^l+1})=0$, then
      \begin{equation*}
      N_{\rho}(a,b)=
      \begin{cases}
      p^{2e-3}+p^{e+m-2}, & \mbox{when }\rho = b; \\
      p^{2e-3}, & \mbox{when } \rho \neq b.
    \end{cases}
    \end{equation*}
      \item[(ii)] $Tr(\gamma^{p^l+1}) \neq 0$, then
      \begin{equation*}
      N_{\rho}(a,b)=
      \begin{cases}
      p^{2e-3}-p^{e+m-2}\eta(\rho^2-2b\rho), & \mbox{when } Tr(\gamma^{p^l+1})=\frac{b^2}{4}; \\
      p^{2e-3}-p^{e+m-2}\eta((\rho-b)^2-4Tr(\gamma^{p^l+1})), & \mbox{when } Tr(\gamma^{p^l+1}) \neq \frac{b^2}{4}.
    \end{cases}
    \end{equation*}
    \end{enumerate}
    \item[(2)] $b \in \mathbb{F}_{q}^*/\mathbb{F}_{p}^*$, then $N_{\rho}(a,b)=p^{2e-3}+p^{e+m-3}$.
  \end{enumerate}
\end{enumerate}
\end{lemma}

\textbf{Proof:} We only prove the case 4(1)(ii) here since the proofs of the other cases are very similar.

In this case, we have
$\varphi_{1}=0,$ $\varphi_{2}=0,$ $\varphi_{3}=0$ and
\begin{equation*}
  \varphi_{4}=
  \begin{cases}
    -p^{e+m-3}-p^{e+m-2}\eta(-Tr(\gamma^{p^l+1})), & \mbox{when } \rho = b; \\
    -p^{e+m-3}-p^{e+m-2}\eta((\rho-b)^2)-4Tr(\gamma^{p^l+1})), & \mbox{when } \rho \neq b.
  \end{cases}
\end{equation*}
 from the discussion above. Applying these values into Eq.(5), we can get 
 \begin{equation*}
 \begin{aligned}
      N_{\rho}(a,b)&=p^{2e-3}-p^{e+m-2}\eta((\rho-b)^2-4Tr(\gamma^{p^l+1}))\\&=
      \begin{cases}
      p^{2e-3}-p^{e+m-2}\eta(\rho^2-2b\rho), & \mbox{when } Tr(\gamma^{p^l+1})=\frac{b^2}{4}; \\
      p^{2e-3}-p^{e+m-2}\eta((\rho-b)^2-4Tr(\gamma^{p^l+1})), & \mbox{when } Tr(\gamma^{p^l+1}) \neq \frac{b^2}{4}.\;\;\;\;\;\;\;\;\;\;\;\hfill \square
    \end{cases}
    \end{aligned}
    \end{equation*}

 \begin{lemma}\label{T2}
With the notations above and $m/s \equiv 0\,mod\,2$, we have
\begin{enumerate}
  \item If $a = 0, b = 0$, then $N_{\rho}(a,b)=0$.
  \item If $a = 0, b \neq 0$,
  \begin{enumerate}
    \item[(1)] $b \in \mathbb{F}_{p}^*$, then
    \begin{equation*}
      N_{\rho}(a,b)=
      \begin{cases}
      p^{2e-2}+p^{e+m+s-2}, & \mbox{when } \rho = b; \\
      0, & \mbox{when } \rho \neq b.
    \end{cases}
    \end{equation*}
    \item[(2)] $b \in \mathbb{F}_{q}^*/\mathbb{F}_{p}^*$, then
    $N_{\rho}(a,b)=p^{2e-3}+p^{e+m+s-3}$.
  \end{enumerate}
  \item If $a \neq 0, b = 0$, then
  \begin{enumerate}
    \item[(1)] If the equation $X^{p^{2l}}+X=-a^{p^l}$ has no solution in $\mathbb{F}_{q}$, then $$N_{\rho}(a,b)=p^{2e-3}+p^{e+m+s-3}.$$
    \item[(2)] If the equation $X^{p^{2l}}+X=-a^{p^l}$ has some solution $\gamma$ in $\mathbb{F}_{q}$,
    \begin{equation*}
      N_{\rho}(a,b)=
      \begin{cases}
      p^{2e-3}, & \mbox{when } Tr(\gamma^{p^l+1}) = 0; \\
      p^{2e-3}-p^{e+m+s-2}\eta(\rho^2-4Tr(\gamma^{p^l+1})), & \mbox{when } Tr(\gamma^{p^l+1}) \neq 0.
    \end{cases}
    \end{equation*}
  \end{enumerate}

  \item If $a \neq 0, b \neq 0$,
  \begin{enumerate}
    \item[(1)]  $b \in \mathbb{F}_{p}^*$,
    \begin{enumerate}
      \item[(i)] If the equation $X^{p^{2l}}+X=-a^{p^l}$ has no solution in $\mathbb{F}_{q}$, then $$N_{\rho}(a,b)=p^{2e-3}+p^{e+m+s-3}.$$
      \item[(ii)] If the equation $X^{p^{2l}}+X=-a^{p^l}$ has some solution $\gamma$ in $\mathbb{F}_{q}$, and $Tr(\gamma^{p^l+1})=0$, then
      \begin{equation*}
      N_{\rho}(a,b)=
      \begin{cases}
      p^{2e-3}+p^{e+m+s-2}, & \mbox{when }\rho = b; \\
      p^{2e-3}, & \mbox{when } \rho \neq b.
    \end{cases}
    \end{equation*}
      \item[(iii)] If the equation $X^{p^{2l}}+X=-a^{p^l}$ has some solution $\gamma$ in $\mathbb{F}_{q}$, and  $Tr(\gamma^{p^l+1}) \neq 0$, then
      \begin{equation*}
      N_{\rho}(a,b)=
      \begin{cases}
      p^{2e-3}-p^{e+m+s-2}\eta(\rho^2-2b\rho), & \mbox{when } Tr(\gamma^{p^l+1})=\frac{b^2}{4}; \\
      p^{2e-3}-p^{e+m+s-2}\eta((\rho-b)^2-4Tr(\gamma^{p^l+1})), & \mbox{when } Tr(\gamma^{p^l+1}) \neq \frac{b^2}{4}.
    \end{cases}
    \end{equation*}
    \end{enumerate}
    \item[(2)] $b \in \mathbb{F}_{q}^*/\mathbb{F}_{p}^*$, then $N_{\rho}(a,b)=p^{2e-3}+p^{e+m+s-3}$.
  \end{enumerate}
\end{enumerate}
\end{lemma}

\textbf{Proof:} By a similar argument as that of Lemma \ref{T1}, we can get the desired conclusions. So we will not discuss it any more.\hfill $\square$

\textbf{Proof of Theorem \ref{weight1}:} With Lemma \ref{quadratic character}, Lemma \ref{T1} and Eq.(4), we can get the weights of the codewords, which are shown in Table 1. We denote the non-zero weights of the lines $1-5$ in Table \ref{1} by $wt_i$, and the corresponding multiplicity by $A_{wt_i}(1\leq i\leq 5)$.

We find that if  $a=0$ and $b\in\mathbb{F}_p^*$, $wt_{1}=p^{2e-2}+p^{e+m-2}$, and thus, $A_{wt_1}=p-1;$ if $a\in\mathbb{F}_q$ and $b\in\mathbb{F}_q^*\backslash\mathbb{F}_p^*$,  $wt_{2}=(p-1)(p^{2e-3}+p^{e+m-3}),$ and so $A_{wt_2}=p^{e}(p^{e}-p).$ Then by the first three Pless Power Moments \cite{HP}, we have
\begin{eqnarray*}
&&\sum_{i=1}^5 {A_{wt_i}}=p^{2e}-1,\qquad \qquad \qquad \qquad \qquad \quad \\
&&\sum_{i=1}^5 {wt_i}A_{wt_i}=p^{2e - 1}(p - 1)(p^{2e - 2}+p^{e+m-2}) , \\
&&\sum_{i=1}^5 {wt_i^2A_{wt_i}}= p^{2e - 2}(p - 1)(p^{2e - 2} +p^{e+m -2})(p^{2e - 1}+p^{e+m -1}-p^{2e-2}-p^{e+m -2}+1).
\end{eqnarray*}
Using Maple 18, we can get the values of $A_{wt_{3}}$-$A_{wt_{5}}$, which are given in Table 1.
Thus we can get the complete weight enumerator presented in Theorem \ref{weight1}, and complete the proof.\hfill $\square$

\textbf{Proof of Theorem \ref{weight2}:} With Lemma \ref{quadratic character}, Lemma \ref{T2} and Eq.(4), we can get the weights of the codewords, which are shown in Table 2. We denote the non-zero weights of the lines $1-5$ in Table \ref{2} by $wt_i$, and the corresponding multiplicity by $A_{wt_i}(1\leq i\leq 5)$.

We find that if  $a=0$ and $b\in\mathbb{F}_p^*$, $wt_{1}=p^{2e-2}+p^{e+m+s-2}$, and thus, $A_{wt_1}=p-1;$ if $a\in\mathbb{F}_q$ and $b\in\mathbb{F}_q^*\backslash\mathbb{F}_p^*$, $wt_{2}=(p-1)(p^{2e-3}+p^{e+m+s-3}),$ and the multiplicity $A_{wt_{21}}=p^{e}(p^{e}-p);$ if $b \in \mathbb{F}_p$ and $X^{p^{2l}}+X=-a^{p^{l}}$ has no solution in $\mathbb{F}_q$, by Lemma \ref{weil sums4}, the multiplicity $A_{wt_{22}}=p(p^e-p^{e-2s})$; thus $A_{wt_2}=A_{wt_{21}}+A_{wt_{22}}=p^e(p^e-p^{1-2s})$. Then by the first three Pless Power Moments \cite{HP}, we have
\begin{eqnarray*}
&&\sum_{i=1}^5 {A_{wt_i}}=p^{2e}-1,\qquad \qquad \qquad \qquad \qquad \quad \\
&&\sum_{i=1}^5 {wt_i}A_{wt_i}=p^{2e - 1}(p - 1)(p^{2e - 2}+p^{e+m+s-2}) , \\
&&\sum_{i=1}^5 {wt_i^2A_{wt_i}}= p^{2e - 2}(p - 1)(p^{2e - 2} +p^{e+m+s -2})(p^{2e - 1}+p^{e+m+s -1}-p^{2e-2}-p^{e+m+s -2}+1).
\end{eqnarray*}
Using Maple 18, we can get the values of $A_{wt_{3}}$-$A_{wt_{5}}$, which are given in Table 2. Thus we can get the complete weight enumerator presented in Theorem \ref{weight2}, and complete the proof.\hfill $\square$

\section{Concluding remarks}\label{section-5}\rm
In this paper, inspired by the work in \cite{XXRF}, a class of four-weight or five-weight linear codes were constructed with their complete weight enumerators settled using Weil sums. At the same time, some optimal or almost optimal linear code was found. It would be nice if more linear codes with few weights can be presented.


\end{document}